\documentclass[preprint,aps,12pt,preprintnumbers,eqsecnum,nofootinbib]{revtex4}
\usepackage{graphicx}
\usepackage{subfigure}

\usepackage{color}
\usepackage{amssymb,amsmath}

\newcommand{\beq}{\begin{equation}}
\newcommand{\enq}{\end{equation}}

\begin{document}
%
\title{\vspace*{0.5in} 
Flavor-Nonuniversal Dark Gauge Bosons \\ and the Muon $g-2$
\vskip 0.1in}
\author{Christopher D. Carone}\email[]{cdcaro@wm.edu}

\affiliation{High Energy Theory Group, Department of Physics,
College of William and Mary, Williamsburg, VA 23187-8795}
\date{January 2013}
\begin{abstract}
The possibility of explaining the current value of the muon anomalous magnetic moment
in models with an additional U(1) gauge symmetry that has kinetic mixing with hypercharge are 
increasingly constrained by dark photon searches at electron accelerators.  Here we present a scenario in 
which the couplings of new, light gauge bosons to standard model leptons are naturally weak and flavor 
non-universal.  A vector-like state that mixes with standard model leptons serves as a portal between the 
dark and standard model sectors.  The flavor symmetry of the model assures that
the induced couplings of the new gauge sector to leptons of the first generation are very small and lepton-flavor-violating 
processes are adequately suppressed.  The model provides a framework for constructing ultraviolet 
complete theories in which new, light gauge fields couple weakly and dominantly to leptons of the 
second or third generations.
\end{abstract}
\maketitle

\section{Introduction} \label{sec:intro}
In recent years, the energy, intensity and cosmic frontiers have come into focus as the three 
domains in which future discoveries in elementary particle physics are likely to arise.  Among the 
possibilities that might be revealed at the intensity frontier are ``dark sectors"~\cite{dark}, with  
particles much lighter than the electroweak scale that have remained undetected due to their 
very weak interactions with standard model particles.  The gravitational effects of dark matter 
is indirect evidence that at least one dark particle exists. The possibility that dark matter may 
also couple to spontaneously broken dark gauge forces has been motivated by the observation 
that sub-GeV dark gauge boson decays could explain the electron and positron excesses observed in
the cosmic ray spectra~\cite{todm}.  More generally, sectors that are only weakly coupled to the standard 
model arise quite generically in string theory~\cite{string}, so that investigations of other possible dark
sectors~\cite{genlit} continue to be well motivated.

Given that the particle content and gauge structure of a generic dark sector is not well constrained, the largest number of analyses have 
assumed the simplest construction, a dark U(1) gauge field that has a small kinetic mixing with hypercharge~\cite{kU1}.  Interestingly, it has 
been pointed out that a discrepancy between the observed value of the muon anomalous magnetic moment and the expectation of the 
standard model can be explained in this scenario~\cite{kU1gm2}.   However, the region of the mass-kinetic mixing plane in which this 
result can be achieved is very rapidly being excluded by bounds from dark photon searches~\cite{dps}, in particular, the most recent results 
from the Mainz Microtron experiment~\cite{mainz}.  Dominant kinetic mixing will always assure that a dark photon couples identically to 
electrons and muons, so there is no way to escape this bound.  The possibility that dark forces might couple differently to muons than electrons 
has been discussed in a somewhat different context, {\em i.e.}, the proton charge radius problem, in Ref.~\cite{pospelov}, but from an 
effective field theory perspective.  Here we will present a complete theory in which such a possibility is naturally realized.

Although an explanation of the muon anomalous magnetic moment discrepancy is certainly worthwhile, it is arguably more important that the present 
work shows, in a complete theory, how a dark sector can naturally couple both weakly and preferentially to charged leptons of the second or third generation.   
This may provide motivation for a more intensive phenomenological consideration of processes involving muons and taus as a means of discerning couplings 
to new dark sector particles.  The basic structure of the model is this: the dark gauge and higgs particles couple directly to a vector-like charged lepton that has 
gauge quantum numbers that allow mixing with the right-handed electron, muon or tau.   However, the lepton sector is also constrained by a discrete, $A_4$ flavor 
symmetry that is the origin of neutrino tribimaximal mixing.  This symmetry leads to preferential mixing between the vector-like state and only one standard model lepton 
flavor, while simultaneously keeping lepton-flavor-violating effects suppressed.  The couplings between the standard model and the dark sector particles are naturally 
small since they are proportional to a small mixing angle that is set by the ratio of the  dark to vector-like mass scales.  Lepton-flavor-universal couplings due to kinetic 
mixing with hypercharge are loop suppressed, since the dark sector gauge group in the proposed model is non-Abelian.

Our letter is organized as follows.  In the next section, we describe the basic structure of the 
theory and show that the muon anomalous magnetic moment discrepancy can be resolved.  In 
Sec.~\ref{sec:flavor}, we explain how the  pattern of couplings assumed in Sec.~\ref{sec:port} can be achieved 
via the discrete flavor symmetry $A_4$.  We briefly discuss other phenomenological 
bounds in Sec.~\ref{sec:other}, and in Sec.~\ref{sec:concl} we conclude.

\section{The Portal} \label{sec:port}
The gauge group of the model is $G_{{\rm SM}} \times$SU($2$)$_D$, where $G_{{\rm SM}}$ is the standard model (SM) gauge group and the additional factor represents the gauge group of the dark sector. Since the dark gauge group is non-Abelian, kinetic mixing with hypercharge does not appear immediately through a renormalizable term in the Lagrangian.   We assume the existence of a vector-like doublet 
\begin{equation}
E_L^a \, , \,\,\,\,\, E_R^a\, ,
\end{equation}
where $a$ is an SU($2$)$_D$ index.  These fields have the same $G_{{\rm SM}}$ quantum numbers as a right-handed electron, muon or tau.  Note that the theory is free of anomalies, since the new fermions come in a vector-like pair.  Since the $E$ fields carry both SM and dark quantum numbers, they serve as a portal between the SM and dark sectors.

We assume that the dark gauge symmetry is completely broken by the vacuum expectation value (vev) of a complex scalar doublet $\phi^a$,
\begin{equation}
\langle \phi \rangle = \left(\begin{array}{c} 0 \\ v_D/\sqrt{2} \end{array} \right) \,.
\end{equation}
A renormalizable coupling between $\phi$ and the SM Higgs field $H$, namely $\phi^\dagger\phi H^\dagger H$, can serve as another portal between the SM and dark sectors.  However, to simplify the model parameter space in the present analysis, we will only consider the case in which this coupling is
negligibly small. This limit is consistent with LHC data on the recently discovered $125$~GeV boson which (perhaps aside from the decay rate 
to two photons) shows no substantial deviations from the expectations for a standard model Higgs boson. 

Mass mixing between the $E$ and SM lepton fields can occur when the SU($2$)$_D$ gauge symmetry is broken.   This mixing will induce a coupling between the dark gauge boson multiplet and ordinary SM leptons that is suppressed by a small mixing angle, one proportional to $v_D/v$, where $v=246$~GeV is the electroweak scale.  Like most dark photon models, we will typically assume that 
the dark gauge bosons have masses less than $1$~GeV.   Mass mixing originates through the Yukawa couplings
\begin{equation}
{\cal L} \supset h \, \overline{\mu}_R \phi^*_a E^a_L + h' \, \epsilon_{ab}\, \overline{\mu}_R \phi^a E^b_L 
+ \mbox{ h.c. } \, ,
\label{eq:mummix}
\end{equation}
Here we have taken the SM lepton field to be of the second generation.  Generically one expects similar couplings 
involving the $e$ and $\tau$.  Our model, however, is not generic.  We will show in Sec.~\ref{sec:flavor} that the flavor 
structure of Eq.~(\ref{eq:mummix}) follows accurately from the flavor symmetries of the model. For the remainder of 
this section, we will simply assume the mixing given in Eq.~(\ref{eq:mummix}) and study the new contributions to the 
muon $g-2$.

In order to specify the model, one must know the vector-like lepton mass $M$, the mass and gauge coupling of the dark gauge bosons, $m_V$ and $g_D$ respectively, and the mass of the dark higgs, $m_\varphi$. (The dark higgs is the physical fluctuation about $v_D$ in the doublet field $\phi$.)  In addition, one must know the mixing angles that are functions of the couplings $h$ and $h'$ in Eq.~(\ref{eq:mummix}).  While a complete exploration 
of this six-dimensional parameter space is impractical, the essential physics can be illustrated by adopting a few simplifying assumptions.  We take $m_V=m_\varphi$, so that a single mass scale characterizes the dark sector.  In addition, we set the coupling $h'=0$, which renders the analysis of mixing between the SM leptons and the vector-like states more transparent. (This latter assumption can be elevated to a definition of the model if one imposes a discrete symmetry that forbids the $h'$ coupling.  A $Z_3$ symmetry in which $E_{L,R} \rightarrow \omega E_{L,R}$ and $\phi^a \rightarrow \omega \, \phi^a$, with $\omega^3=1$, is a suitable example of such a symmetry.)
\begin{figure}[t]
    \centering
    \includegraphics[width=8cm,angle=0]{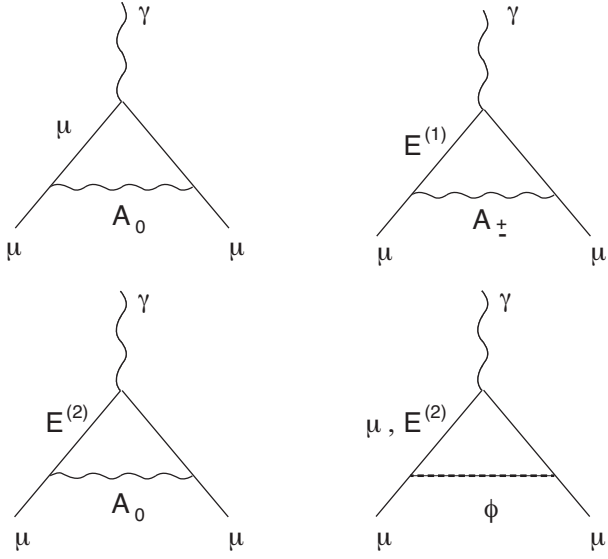}
    \caption{New contributions to the muon anomalous magnetic moment. 
    \label{fig:gm2}}
\end{figure}

Given these assumptions, mass mixing between the SM and vector-like states are given
by a two-by-two mass matrix,
\begin{equation}
{\cal L} \supset \overline{\Psi}_L
\left(\begin{array}{cc} m_\mu & 0 \\ \delta m & M \end{array} \right) \Psi_R+ \mbox{h.c.} \, ,
\end{equation}
where $\Psi = (\mu \, , \, E^{(2)})^T$ and $\delta m = h\,v_d / \sqrt{2}$.   The relationship between the gauge and mass eigenstate bases is given by
\begin{equation}
\Psi_{j} = \left(\begin{array}{cc} c_{j} & s_{j} \\ \!\!-s_{j} & c_{j}
\end{array}\right) {\Psi_{j}}^{\rm mass} \,
\end{equation}
where $s_j = \sin\theta_j$, $c_j=\cos\theta_j$, and $j$ is either $L$ or $R$.  With
$\delta m \ll M$, the mixing angles are well approximated by 
$\theta_R \approx \delta m/M$ and $\theta_L \approx (m_\mu/M)(\delta m / M)$.
Due to this mixing, the SU($2$)$_D$ gauge bosons, which otherwise would couple
only to the vector-like states, now also couple to standard model leptons.  In terms of
mass eigenstate fields, these
couplings include
\begin{eqnarray}
{\cal L} &\supset& 
[-\frac{1}{2} s_R^2 \, g_D \, (\overline{\mu}_R \!\not\!\!{A_0} \mu_R)
-\frac{1}{\sqrt{2}} s_R \, g_D \, (\overline{\mu}_R \!\not\!\!{A_-} E^{(1)}_R + \mbox{h.c.})
\nonumber \\
&+& \frac{1}{2} s_R c_R \, g_D \, (\overline{\mu}_R \!\not\!\!{A_0} E^{(2)}_R + \mbox{h.c.})]+ [\mbox{R} \rightarrow \mbox{L}]
\end{eqnarray}
Here we have parameterized the SU($2$) gauge multiplet 
\begin{equation}
A_D^a T^a  = \left(\begin{array}{cc} A_0/2 & A_+/ \sqrt{2} \\ A_-/ \sqrt{2} & -A_0/2 \,,
\end{array}\right)
\end{equation}
where the $T^a$ are the Pauli matrices over two. Note that the $A_0$, $A_+$ and 
$A_-$ are electrically neutral, but the $E^{(1)}$, $E^{(2)}$ and $\mu$ fields have identical couplings to the photon.  Hence there are number of diagrams that contribute to the muon anomalous magnetic moment, as shown in Fig.~\ref{fig:gm2}.

In addition to the gauge-boson exchange diagrams, there are also  diagrams involving exchange of the dark 
Higgs boson $\varphi$, also shown in Fig.~\ref{fig:gm2}.  These follow from Eq.~(\ref{eq:mummix}), which includes the following couplings of the mass eigenstate fields
\begin{eqnarray}
{\cal L} &\supset& \frac{h}{\sqrt{2}} \varphi \left(-s_L c_R \, \overline{\mu}\,\mu + c_L s_R\, \overline{E^{(2)}} E^{(2)}
+[c_L c_R \, \overline{\mu}_R E^{(2)}_L-s_L s_R \,\overline{\mu}_L E^{(2)}_R + \mbox{ h.c.}] \right) .
\end{eqnarray}

\begin{figure}[t]
    \centering
    \includegraphics[width=8cm,angle=0]{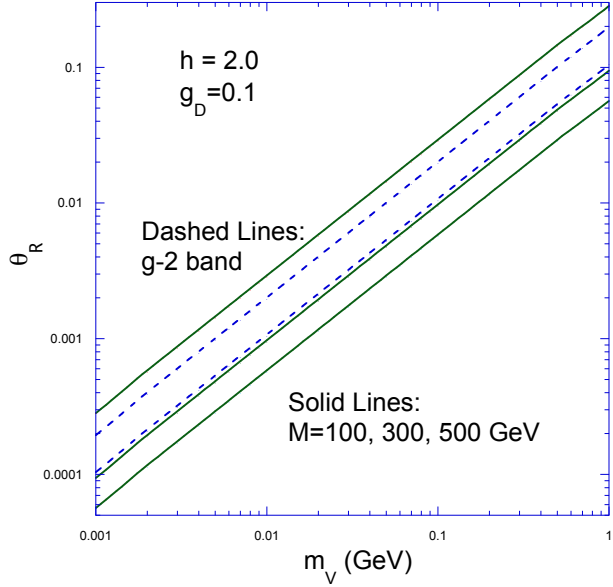}
    \caption{Mass-mixing angle plane, for the values of $g_D$ and $h$ shown. The region between the 
    dashed lines is consistent with the current experimental value of the muon anomalous magnetic 
    moment, at plus or minus two standard deviations.  The solid lines show the vector-like fermion 
    masses, which increase for smaller values of $\theta_R$.
    \label{fig:pspace}}
\end{figure}

The contribution of new fermions and gauge bosons to the muon anomalous magnetic moment have been computed in a general framework in Ref.~\cite{g2ref}, assuming couplings of the form
\begin{equation}
{\cal L} = \overline{\mu} \gamma^\mu (C_V + C_A \gamma^5) F X_\mu + \mbox{h.c.} \, ,
\end{equation}
where $F$ is a generic fermion and $X_\mu$ a generic gauge boson, and
\begin{equation}
{\cal L} = \overline{\mu} \, (C_S + C_P \gamma^5) F S + \mbox{h.c.} \, ,
\end{equation}
where $S$ is a generic scalar.  These results can be applied to the present model given the identifications
shown in Table~\ref{table:one}.
\begin{table}
\begin{tabular}{ccccccc}
$F$  & \qquad & $X_\mu$  & \qquad & $C_V$ & &$C_A$ \\ \hline
$\mu$ && $A_0$ &&$-\frac{1}{4} g_D (s_R^2+s_L^2)$ && $-\frac{1}{4} g_D (s_R^2-s_L^2)$ \\
$E^{(1)}$ && $A_\pm$ && $-\frac{1}{2\sqrt{2}} g_D (s_R + s_L)$ && $-\frac{1}{2\sqrt{2}} g_D (s_R-s_L)$ \\
$E^{(2)}$ && $A_0$ && $\frac{1}{4} g_D (c_R s_R + c_L s_L)$ && $\frac{1}{4} g_D (c_R s_R -c_L s_L)$ \\
\hline \hline \\
$F$ && $S$ && $C_S$ && $C_P$ \\ \hline
$\mu$ && $\varphi$ && $-\frac{1}{\sqrt{2}} h \, (c_R s_L)$ && $0$ \\
$E^{(2)}$ && $\varphi$ && $\frac{1}{2\sqrt{2}} h \, (c_L c_R-s_L s_R)$ && $- \frac{1}{2\sqrt{2}} h \,
(c_L c_R + s_L s_R)$ \\ \hline\hline
\end{tabular}
\caption{Couplings relevant to the anomalous magnetic moment calculation.} \label{table:one}
\end{table}
An exact numerical evaluation of the $g-2$ formulae given in Ref.~\cite{g2ref} have been incorporated 
in Fig.~\ref{fig:pspace}.  Here we show the $m_V$-$\theta_R$ plane, for a fixed choice of the mixing
parameter $h$ and the dark gauge coupling $g_D$, with $m_\varphi = m_V$.  With $m_V$ and $g_D$ fixed, 
the dark scale $v_D$ is determined.   The mass scale of the vector-like sector, $M$, is then fixed by
by its relation to the right-handed mixing angle
\begin{equation}
\theta_R = \left(\frac{\sqrt{2} h}{g_D M} \right) \, m_V \,,
\end{equation}
as indicated by the solid lines in the figure.   The parameters of the model are now fixed and the 
(much smaller) left-handed mixing angle can also be computed.

The region between the dashed lines is where the model accounts for the discrepancy between 
the experimental and SM values of the muon $(g-2)/2$~\cite{rpp}
\begin{equation}
\Delta a_\mu = a_\mu^{\rm exp} - a_\mu^{\rm SM} = 287 \pm 80 \times 10^{-11} \, ,
\end{equation}
allowing two-standard deviation variation about the central value of $\Delta a_\mu$.  The solid lines represent
vector-like fermion mass contours, for $M=100$, $300$ and $500$~GeV.  Collider searches
for heavy charged leptons require only that $M>100.8$~GeV at the 95\%C.L.~\cite{rpp}.  The
solid and dashed bands still overlap for $h$ closer to $1$, but for values of $M$ closer
to its experimental lower bound.

\section{The Flavor Symmetry}\label{sec:flavor}

There is a vast literature on flavor symmetries that can account for the very non-generic pattern of standard model fermion 
masses.   In this section, we argue that the same flavor symmetry that might explain standard model lepton masses can also 
yield the non-generic pattern of couplings assumed in the previous section, with lepton-flavor-violating effects adequately 
suppressed.

A variety of models based on the discrete flavor group $A_4$ have been proposed after it was discovered that $A_4$ symmetry
can provide neutrino mass matrix textures that are consistent with tribimaximal mixing, at lowest order.  Given these promising results, 
it is well motived to consider the consequences of this symmetry in the present model.   The group $A_4$ has a triplet and three singlet 
representations, ${\bf 3}$, ${\bf 1}^0$, ${\bf 1}^+$, and ${\bf 1}^-$.  Typical $A_4$ models assign the left-handed SM leptons to the 
${\bf 3}$ and the right-handed leptons to distinct singlet representations:
\begin{equation}
L_L \sim {\bf 3} \, , e_R\sim {\bf 1}^0 \, , \mu_R \sim {\bf 1}^+ \, , \tau_R \sim {\bf 1}^- \, .
\end{equation}
The standard model Higgs doublet $H$ is a trivial $A_4$ singlet, ${\bf 1}^0$.  For a review of group theory and model building with 
$A_4$ symmetry, we refer the reader to the review in Ref.~\cite{altarelli}.   The symmetry is broken by two flavon fields, $\varphi_S$ and
$\varphi_T$, that both transform as ${\bf 3}$'s:
\begin{equation}
\langle \varphi_S \rangle =\left(\begin{array}{c} v_S \\ v_S \\ v_S \end{array}\right) \, , \,\,\,\,\,\,\,\,\,\,
\langle \varphi_T \rangle =\left(\begin{array}{c} v_T \\ 0 \\  0 \end{array}\right) \,.
\end{equation}
By imposing additional, model-dependent symmetries~\cite{altarelli}, it is arranged that the symmetry breaking in the charged lepton
sector is a consequence of the $\varphi_T$ flavon only, at lowest order, while in the neutrino sector the same statement holds for
the $\varphi_S$ flavon.  Symmetry breaking due to $\varphi_T$ alone leads to a charged lepton mass matrix that is {\em exactly} diagonal,
\begin{equation}
Y_L = \frac{v_T}{\Lambda_F} \left(\begin{array}{ccc} y_e & 0 & 0 \\ 0 & y_\mu & 0 \\ 0 & 0 & y_\tau \end{array}\right) \, ,
\label{eq:yL}
\end{equation} 
where $\Lambda_F$ represents the scale at which flavor physics is generated. The reason for this result is that $\langle \varphi_T \rangle$ breaks 
$A_4$ down to a $Z_3$ under which all off-diagonal elements of Eq.~(\ref{eq:yL}) rotate by a nontrivial phase; hence, these elements remain 
vanishing until this symmetry is broken.   

We now assign $A_4$ quantum numbers to the vector-like states:
 \begin{equation}
 E_L^a \sim E_R^a \sim {\bf 1}^+
 \label{eq:tca}
 \end{equation}
Since this charge assignment is the same as that for $\mu_R$, the mixing terms in Eq.~(\ref{eq:mummix}) are $A_4$ invariant.   However, the
other possible terms are not:
\begin{equation}
\overline{e}_R \phi^*_a E_L^a \sim {\bf 1}^+\, , \,\,\,\,\,\,\,\,\,\,  \overline{\tau}_R \phi^*_a E_L^a \sim {\bf 1}^- \, .
\label{eq:notthere}
\end{equation}
The representations in Eq.~(\ref{eq:notthere}) rotate by a nontrivial phase under the $Z_3$ symmetry that is left unbroken by
$\langle \varphi_T \rangle$, and are forbidden until the $Z_3$ symmetry is broken.  Hence, the same symmetry structure that leads
to a diagonal charged lepton mass matrix when $A_4 \rightarrow Z_3$ assures that the vector-like states mix only with right-handed
muons in the present model. For a different choice in Eq.~(\ref{eq:tca}), one could just as easily arrive at a different model in which mixing with $\tau_R$ is preferred.

We must now consider to what extent our desirable results are spoiled when the $Z_3$ symmetry is broken by the vacuum expectation
value of the other flavon field $\varphi_S$.  Unwanted couplings between $E$ and $e_R$ or $\tau_R$ can occur directly, or 
via Eq~(\ref{eq:mummix}), which is no longer in the mass eigenstate basis if $Y_L$ is non-diagonal.  The value of the $Z_3$-symmetry-breaking
parameter $v_S/\Lambda$, however, is not set by the size of a typical charged lepton Yukawa coupling; it is determined by the mass scale
of the right-handed neutrinos that participate in the see-saw mechanism.    As is standard in $A_4$-based models, we take the right-handed
neutrinos (charge conjugated) to transform as a triplet
\begin{equation}
\nu_R^c \sim {\bf 3} \, .
\end{equation}
We will now show that it is possible to achieve the proper light  neutrino mass scale with $v_S \ll v_T < \Lambda_F$, so that deviations 
from the $Z_3$ symmetric results in the charged SM lepton and vector-like sectors of our model can be made adequately small.  In this 
case, contributions to lepton-flavor-violating processes due to the exchange of the non-standard-model states in the theory will also be 
highly suppressed.  In what follows, we will take the flavor cut off $\Lambda_F$ to be the reduced Planck mass, $M_*$, which provides the
smallest possible values for $v_S/\Lambda_F$.

The Lagrangian terms that lead to a see-saw mechanism in neutrino sector are
\begin{equation}
{\cal L}_\nu = \frac{1}{2}m_R \, \overline{\nu_R^c}\, \nu_R + \frac{1}{2} 
x\,  \overline{\nu_R^c}\, \varphi_S\, \nu_R +[y\,  \overline{L}_L H \nu_R + \mbox{h.c.}] \, ,
\end{equation}
where $x$ and $y$ are dimensionless coupling constants.  This leads to the conventional tribimaximal form for the Dirac and Majorana neutrino mass
matrices~\cite{altarelli} that are input into the see-saw
\begin{equation}
m_{LR} = \left(\begin{array}{ccc} 1 & 0 & 0 \\ 0 & 0 & 1 \\ 0 & 1 & 0 \end{array}\right)\,y \,v  \, , \,\,\,\,\,\,\,\,\,\,
m_{RR} = \left(\begin{array}{ccc} A + 2 B / 3 & -B/3 & -B/3 \\ -B/3 & 2 B/3 & A-B/3 \\ -B/3 & A-B/3 & 2 B/3 \end{array}\right) m_R \, ,
\end{equation}
where $A = 1$, $B=x v_S / m_R$, and $v \approx 246$~GeV is the electroweak scale.  The light neutrino mass matrix then follows from the see-saw formula $m_\nu = m_{LR}^T m_{RR}^{-1} m_{LR}$,
and has the eigenvalues 
\begin{equation}
m_1=\frac{y^2 v^2}{m_R}\frac{1}{(A+B)} \, , \,\,\,\,\,
m_2= \frac{y^2 v^2}{m_R} \frac{1}{A}  \, ,  \,\,\,\,\,
m_3= \frac{y^2 v^2}{m_R} \frac{1}{B-A}  \, .
\label{eq:eigens}
\end{equation}
Neutrino oscillation data indicate the neutrino mass squared splittings $\Delta m_{21}^2 = (7.50 \pm 0.20) \times 10^{-5}$~eV$^2$ 
and $\Delta m_{32}^2 = 2.32^{+0.12}_{-0.08} \times 10^{-3}$~eV$^2$~\cite{rpp}.   These can be accommodated by 
Eq.~(\ref{eq:eigens}) for $y^2 v^2/m_R =0.97 \times10^{-11}$~GeV and $B=x v_S/m_R = 1.198$.  This solution implies  
\begin{equation}
\frac{v_S}{M_*} \approx 3.7 \times 10^{-3} \, \frac{y^2}{x} \, .
\label{eq:vsoms}
\end{equation}
The key point is this: Eq.~(\ref{eq:vsoms}) determines the deviation from the $Z_3$-symmetric limit our theory in which no lepton-flavor-violating 
couplings appear. For ${\cal O}(1)$ values of $x$ and small values of of the neutrino Dirac Yukawa coupling $y$, $Z_3$ symmetry breaking can be 
made extremely small.  It is easy to arrange this from a symmetry perspective by applying, for example, a $Z_N$ flavor symmetry on the right-handed 
neutrinos that allows Majorana but not Dirac mass terms at lowest order.  Since the size of $v_S$ is near the right-handed neutrino mass 
scale, the smallness of the parameter in Eq.~(\ref{eq:vsoms}) is related to how low one can take this scale.  For example, with 
$m_R \sim v_S \sim 20$~TeV, one would have $v_S/M_* \approx  10^{-14}$, with $y$ comparable in size to the electron Yukawa coupling.  This is more than 
adequate to render the theory safe from lepton-flavor-violating effects from the coupling of the dark and standard model sectors.  As an example, loop 
effects will generate an effective operator that contributes to $\ell_i \rightarrow \ell_j \, \gamma$ which is of the form~\cite{altarelli}
\begin{equation}
\frac{1}{\Lambda^2_{\rm NP}} {\overline{L}_L}_i \, H \sigma^{\mu\nu} F_{\mu\nu} Z_{ij} {e_R}_j \,
\end{equation}
where the new physics scale here, $\Lambda_{\rm NP}$,  is related to the vector-like fermion mass scale by
$\Lambda_{\rm NP} \approx 4 \pi M/g_D$, and $Z_{ij}$ is a coupling matrix.  Since $Z$ has the same flavor-structure
as $Y_L$ and we assume an additional symmetry prevents $\varphi_S$ from contributing at lowest order, we expect that
$Z \sim {\cal O} (v_T v_S/\Lambda_F^2)$.  From Eq.~(\ref{eq:yL}), we see that $v_T/\Lambda_F$ can be no smaller than
approximately $0.01$ if the tau Yukawa coupling is to remain perturbative.  So, assuming $\Lambda_F= M_*$ and
$M_R\approx 20$~TeV, we can achieve $Z_{ij} \sim {\cal O} (10^{-16})$.  In contrast, the bound on $Z_{\mu e}$ from
Ref.~\cite{altarelli}, updated to take into account the latest PDG value~\cite{rpp} for the $\mu \rightarrow e  \gamma$ branching fraction,
gives $Z_{\mu e} < 4.5 \times 10^{-9} [\Lambda_{\rm NP}/{\rm 1 TeV} ]^2$,
which implies $Z_{\mu e}< 7.1 \times 10^{-7}$ for $g_D=0.1$ and $M=100$~GeV.   This bound is easily satisfied.  Similar conclusions can be drawn
from other possible lepton-flavor-violating processes, which we will not consider in explicit detail here.  

\section{Other phenomenological issues} \label{sec:other}

Before concluding, we comment on a few of the other potential bounds on the model. It has been 
noted~\cite{barger} in the context of other models~\cite{pospelov} in which a right-handed muon 
couples to a new gauge field $V_\mu$ that the effective coupling $g_R \,\overline{\mu}_R\!\!\!\not \!V \mu_R$ 
is bounded by the absence of missing mass events in the leptonic kaon decay 
$K \rightarrow \mu X$.  In the present model, the magnitude of $g_R$ is $s_R^2 g_D/2 $, which ranges in Fig.~\ref{fig:pspace}
from $10^{-8}$ to $10^{-4}$ between approximately $5$ to $300$~MeV.  We comment on this mass range 
since it roughly corresponds the to the one over which limits on $g_R$ as a function of vector boson 
mass are given in Fig.~3 of Ref.~\cite{barger};  the limit is always larger than $10^{-3}$. Hence, the 
model parameter space is not severely constrained based on this consideration.   

The bounds on the dark sector states from accelerator experiments~\cite{bjork} depend on how these states 
decay, and their are many possible scenarios. Here we discuss one interesting case that corresponds 
to the largest parameter region shown in Fig.~\ref{fig:pspace}: if $m_V < 2 m_\mu$, the dark gauge 
fields will decay only to electrons via loop-suppressed kinetic mixing and we can compare 
to the bounds from dark photon searches.  The one-loop diagrams that lead to mixing between the 
the dark gauge bosons and hypercharge lead to effective operators of the form
\begin{equation}
{\cal L}_{\rm eff} = \frac{c_1}{M^2} \phi^\dagger F_D^{\mu\nu} \phi {F_Y}_{\mu\nu}
+\frac{c_2}{M^2} \tilde{\phi}^\dagger F_D^{\mu\nu} \tilde{\phi} {F_Y}_{\mu\nu}
+[\frac{c_3}{M^2} \tilde{\phi}^\dagger F_D^{\mu\nu} \phi {F_Y}_{\mu\nu} + \mbox{h.c.}] \, ,
\label{eq:loopkm}
\end{equation}
where $\tilde{\phi} = \epsilon^{ab} \phi^*_b$. The diagrams that generate these operates involve the vector-like fermions (and hence the 
suppression by $M^2$) as well as two vertices from the the interaction terms in Eq.~(\ref{eq:mummix}).   The couplings $c_1$, $c_2$, and $c_3$ 
are of order $h^2 g_D g_Y/(16 \pi^2)$, $h h' g_D g_Y/(16 \pi^2)$, and ${h'}^2 g_D g_Y/(16 \pi^2)$, respectively.   Kinetic mixing is conventionally 
parameterized as $\chi F_i^{\mu\nu} {F_Y}_{\mu\nu}/2$, where $F_i$ is the field strength tensor corresponding to the dark gauge field $A_i$.  From 
the first term in Eq.~(\ref{eq:loopkm}), for example, we estimate
\begin{equation}
\chi \sim \frac{1}{4 \pi^2} g_Y \frac{h^2}{g_D} \left(\frac{m_V}{M}\right)^2 \,.
\end{equation}
The most constraining beam dump bound~\cite{bdbnds} for small kinetic mixing is from the SLAC E137 experiment.  However, the E137 bound does 
not exclude $\chi$ below $\sim 4 \times 10^{-8}$.   For the parameter values $h=2$ and $g=0.1$, used in Fig.~\ref{fig:pspace}, and $M=200$~GeV, we 
find that $\chi$ is smaller than this value for $m_V \alt 67$~MeV.  For $m_V \agt 210$~MeV, E137 gives no further bounds.  Of course, the dark 
sector could include dark matter candidates into which the the dark gauge and Higgs fields decay.  In this case, dark photon searches at 
accelerators would present no further bounds on the model.

\section{Conclusions} \label{sec:concl}

In this letter, we have shown how one can construct a dark sector that couples preferentially to a 
single flavor of standard model charged leptons, without generating large lepton-flavor-violating effects. The 
dark gauge and higgs bosons couple to ${\cal O}(100)$~GeV vector-like fermions that have the correct gauge quantum numbers to 
mix with either the right-handed electron, muon or tau.   However, the lepton sector is also constrained by an $A_4$ flavor symmetry 
that is the origin of neutrino tribimaximal mixing.  This symmetry assures that the vector-like state mixes with only one flavor of charged lepton while the charged lepton Yukawa matrix remains very accurately diagonal.  Deviations from this flavor 
structure are related to the ratio of the right-handed neutrino mass scale to the cut off of the effective 
theory, and can be made adequately small.  Moreover, lepton-flavor-universal couplings via kinetic mixing 
are loop suppressed since the dark sector gauge group is non-Abelian.

If the vector-like states mix with leptons of the second generation (which has been the assumption in this letter) contributions from the 
non-standard-model particles can account for the current deviation of the muon anomalous magnetic moment from the standard model expectation.  More 
importantly, however, the construction described in this letter can be used to create ultraviolet complete theories (at least up to the high scale at which 
flavor is generated) in which dark sector gauge bosons couple preferentially to either muons or taus.  This provides motivation for exploring a vast realm 
of phenomenological processes involving second and third generation leptons as another possible window on a dark sector.

\begin{acknowledgments}  
This work was supported by the NSF under Grant PHY-1068008.  In addition, the author thanks Joseph J. Plumeri II for his generous support.
\end{acknowledgments}


\end{document}